\documentclass[twoside,reqno]{amsart}

\usepackage{epsfig}
\usepackage{amsmath,amsthm,amsfonts,amssymb,amscd,euscript}

\graphicspath{ {Figures/} }

\input{epsf}

\newcommand{\filebegin}{\begin{document}}
\newcommand{\fileend}{\end{document}}
\makeatother
\newcommand{\lo}{\longrightarrow}
\newcommand{\NMM}{\hspace*{2mm}}

\renewcommand{\baselinestretch}{1.1}

\input{epsf}
\renewcommand{\baselinestretch}{1.1}
\def\n{\noindent}

\numberwithin{equation}{section}
\def\mapdown#1{\Big\downarrow\rlap
{$\vcenter{\hbox{$\scriptstyle#1$}}$}}
\newtheorem{theorem}{Theorem}[section]
\newtheorem{lemma}[theorem]{Lemma}
\newtheorem{proposition}[theorem]{Proposition}
\newtheorem{corollary}[theorem]{Corollary}

\theoremstyle{definition}
\newtheorem{definition}[theorem]{Definition}
\newtheorem{example}[theorem]{\sc Example}
\newtheorem{xca}[theorem]{Exercise}
\theoremstyle{remark}
\newtheorem{remark}[theorem]{Remark}

\begin{document}

\setcounter{page}{1} \noindent
Iranian Journal of Mathematical Sciences and Informatics \\
Vol. x, No. x (xxxx), pp xx-xx \\
DOI: xx.xxxx/ijmsi.20xx.xx.xxx

\vspace*{2cm}
\begin{center}
{\bf\large Single-Point Visibility Constraint Minimum Link Paths in Simple Polygons}
 \\[0.5cm]
{Mohammad Reza Zarrabi$^a$, Nasrollah Moghaddam Charkari$^{a,*}$
\let\thefootnote\relax\footnotetext{$^*$Corresponding Author\\[0.2cm]
Received September 2018; Accepted January 2019 Academic Center for Education, Culture and Research TMU} \\[2mm]
$^a$Faculty of Electrical Engineering and Computer Science, Tarbiat Modares  University, Tehran, Iran\\[2mm]
{\tt E-mail: m.zarabi@modares.ac.ir}\\
{\tt E-mail: charkari@modares.ac.ir}
} \\[2mm]
\end{center}
\vspace*{0.5cm}

\begin{quotation}
\noindent
{\footnotesize {\sc Abstract.} We address the following problem: Given a simple polygon $P$ with $n$ vertices and two points $s$ and $t$ inside it, find a minimum link path between them such that a given target point $q$ is visible
from at least one point on the path.
The method is based on partitioning a portion of $P$ into a number of faces of equal link distance from a source point. This partitioning is essentially a shortest path map (SPM). In this paper, we present an optimal algorithm with $O(n)$ time bound, which is the same as the time complexity of the standard minimum link paths problem.
}
\end{quotation}
\
\\[0.5cm]

\markboth
{M. R. Zarrabi, N. M. Charkari}
{Single-Point Visibility Constraint Minimum Link Paths in Simple Polygons}

\section{Introduction}
Link paths problems in computational geometry have received considerable attention in recent years due to not only their theoretical beauty, but also their wide range of applications in many areas of the real world.
A minimum link path is a polygonal path between two points $s$ and $t$ inside a simple polygon $P$ with $n$ vertices that has the minimum number of links.
Minimum link paths are fundamentally different from traditional Euclidean shortest path,
which has the shortest length among all the polygonal paths without crossing edges of $P$.
Minimum link paths have important applications in various areas like robotic, motion planning,
wireless communications, geographic information systems, image processing, etc.
These applications benefit from minimum link paths since turns are costly while straight line movements are inexpensive.

Many algorithms structured around the notion of link path have been
devised to parallel those designed using the Euclidean path.
In \cite {Suri_1986}, Suri introduced a linear time algorithm for computing a minimum link path between two points inside a triangulated simple polygon $P$
(Ghosh presented a simpler algorithm in \cite {Ghosh_1991}).
After that, Suri developed the proposed solution to a query version by constructing a window partition in linear time for a fixed point inside $P$ \cite {Suri_1990}.
This window partition is essentially a shortest path map, because it divides the simple polygon into
faces of equal link distance from a fixed source point.
By contrast, the work of Arkin et al. \cite {Arkin_1995} supports $O(\log n)$ time queries between
any two points inside $P$ after building $n$ shortest path maps for all vertices of $P$, i.e., the total time complexity for this construction is $O(n^2)$
(an optimal algorithm for this case was presented by Chiang et al. in \cite {Chiang_1997}).
On the other hand, when there are holes in the polygon, Mitchell et al. \cite {Mitchell_1992} proposed an
incremental algorithm with $O(E\alpha(n)\log^2 n))$ time bound, where
$n$ is the total number of edges of the obstacles, $E$ is the size of the visibility graph, and $\alpha(n)$ denotes
the extremely slowly growing inverse of Ackermann's function.
An interesting survey of minimum link paths appears in \cite {Maheshwari_2000}.

Minimum link paths problems are usually more difficult to solve than equivalent Euclidean shortest path problems since optimal paths that are unique under the Euclidean metric need not be unique under the link distance. Another difficulty is that, Euclidean shortest paths only turn at reflex vertices of $P$ while minimum link paths can turn anywhere.

The problem is studied with several variations.
One of these variations is to constrain the path to view a point $q$ from at least one point on the path.
One possible application is resource collection in which a robot moving from a source point to a destination one has to collect some resources found in a certain region.
Some other applications are visibility related constraints such as wireless communication or guarding systems, where direct visibility is crucial.
In this paper, we study the constrained version of minimum link paths problem based on the shortest path map called SPM approach. The proposed algorithm runs in linear time.

Section 2 introduces the basic definitions and lemmas.
Section 3 presents our algorithm for a visibility point $q$.
We conclude in Section 4 with some open problems.

\section{Preliminaries}

\begin{figure}
	\centering
    \includegraphics[width=5.5cm,keepaspectratio=true]{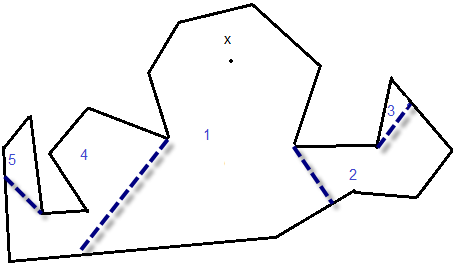}
	\caption{$\mathrm{SPM(x)}$ and its faces}
	\label{fig:Faces}
\end{figure}

\begin{figure}
	\centering
    \includegraphics[width=4cm,keepaspectratio=true]{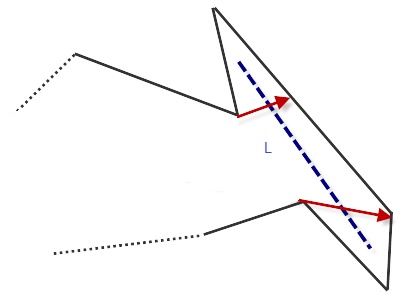}
	\caption{Dividing any line segment $L$ to at most three parts}
	\label{fig:Lemma}
\end{figure}

We use the following notation throughout the paper:

\begin{enumerate}
\item  $V(x):$ the visibility polygon of a point $x$ $\in$ $P$
\item  $\pi_L(x,y):$ a minimum link path between two points $x$ and $y$ in $P$
\item  $D_L(x,y):$ the link distance (minimum number of links) of $\pi_L(x,y)$
\item  $D_\pi(x,y):$ the number of links between $x$ and $y$ on the path $\pi$
\item  $Pocket(x):$ the regions of $P$, invisible to $x$
\end{enumerate}

\begin{definition}
\textbf {q-visible path:}
A minimum link path, which has a non-empty intersection
with $V(q)$ for any point $q$ $\in$ $P$.
\end{definition}

\begin{definition}
\textbf {SPM(x):}
Let $x$ be a point in $P$. The edges of $V(x)$ either are (part of) edges
of $P$ or chords of $P$. The edges of the latter variety are
called windows of $V(x)$. The set of points at link distance one from
$x$ is precisely $V(x)$. The points of link distance two are the points in
$P-V(x)$ that are visible from some windows of $V(x)$. Repeatedly,
applying this procedure partitions $P$ into faces of constant link
distance from $x$ \cite {Suri_1990}.
For the sake of consistency,
we call this partitioning the shortest path map of $P$ with respect to $x$ (see Figure~\ref{fig:Faces}).
\end{definition}

The following two lemmas are the fundamental properties of SPM used in our algorithm:

\begin{lemma}\label{Linear}
For any point $x$ $\in$ $P$, $SPM(x)$ is constructed in $O(n)$ time \cite {Suri_1990}.
\end{lemma}

\begin{lemma}\label{Intersection}
Given a point $x$ $\in$ $P$,
any line segment $L$ intersects at most three faces of $SPM(x)$ \cite {Arkin_1995}.
\end{lemma}

Figure~\ref{fig:Lemma}
illustrates the worst case intersection described in Lemma~\ref{Intersection}.
Our goal is to compute a $q$-$visible$ path inside a simple polygon $P$ with $n$ vertices.

\section{The Algorithm}

\begin{figure}
	\centering
    \includegraphics[width=5cm,keepaspectratio=true]{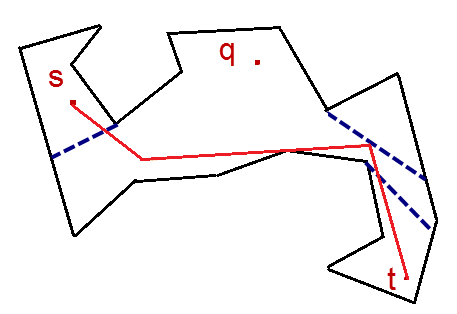}
	\caption{Step 4 of the algorithm}
	\label{fig:Pockets}
\end{figure}

Suppose that three points $s$, $t$ and $q$ are given in $P$.
We sketch the generic algorithm for computing a $q$-$visible$ path between $s$ and $t$ as follows:

\begin{enumerate}
\item If $D_L(s,q)=1$ or $D_L(t,q)=1$, report $\pi_L(s,t)$.

\item Compute $V(q)$ and $Pocket(q)$.

\item Use the point location algorithm for $s$ and $t$ to
      determine whether they are in the same region in $Pocket(q)$ or not.

\item If $s$ and $t$ are in two different regions of $Pocket(q)$, report $\pi_L(s,t)$.
      Since this path crosses $V(q)$,
      it is a $q$-$visible$ path (see Figure~\ref{fig:Pockets}).

\item Suppose that $e$ is a chord of $P$ that separates $V(q)$ and the region of $Pocket(q)$, which contains both
      $s$ and $t$. Indeed, $e$ is a window of $V(q)$ and divides $P$ into two subpolygons, only
      one of which contains $q$. We define $p$ as the subpolygon containing $q$.
      A $q$-$visible$ path should have a non-empty intersection with $p$ (note that $e \in p$).

\item Compute both $\mathrm{SPM(s)}$ and $\mathrm{SPM(t)}$.
      Add two labels $S_j$ and $T_k$ to each \emph{face} of $\mathrm{SPM(s)}$ and $\mathrm{SPM(t)}$
      as the link distance from $s$ and $t$ to those faces, respectively
      ($1 \leq j \leq ||\mathrm{SPM(s)}||$, $1 \leq k \leq ||\mathrm{SPM(t)}||$ and $||.||$ denotes the number of faces
      of $\mathrm{SPM}$).

\item Use the map overlay technique to find the intersections of $\mathrm{SPM(s)}$, $\mathrm{SPM(t)}$ and $p$.
      Construct the planar subdivision of $p$ with new faces (called \emph{Cells}) by
      the \emph{quad view} data structure \cite {Finke_1995}.
      Use a set $Ce$ as a reference to these \emph{Cells} (each element of this set points to the
      value or structure of a \emph{Cell}).

\item Assign the value of each \emph{Cell}, i.e., $Ce(i)= s_i + t_i$, where $1 \leq i \leq$ the number of \emph{Cells} and
      $s_i=S_j$, $t_i=T_k$.
      In other words, $s_i$ and $t_i$ are the link distances from any point $x \in Ce(i)$ to $s$ and $t$, respectively.

\item Find the minimum value of $Ce(i)$ for some $i$.
      It ensures that a $q$-$visible$ path between $s$ and $t$ enters $Ce(i)$.

\item Report $\pi_L(s,x)$ appended by $\pi_L(x,t)$, where $x$ is a point in $Ce(i)$.
      The link distance of this new path would be $s_i + t_i$ (Lemma~\ref{Side}).
\end{enumerate}

\begin{figure}
	\centering
    \includegraphics[width=9.5cm,keepaspectratio=true]{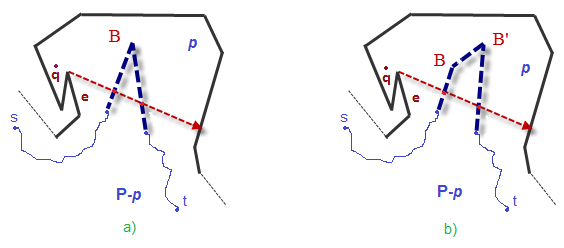}
	\caption{Two points $s$ and $t$ lie on the same side of $e$}
	\label{fig:Oneside}
\end{figure}

\begin{lemma}\label{Side}
Assume $s$ and $t$ lie on the same side of $e$. Then, the link distance of a $q$-$visible$ path
$\pi$ between $s$ and $t$ is: $s_i+t_i$ ($i$ might not be unique).
\end{lemma}

\begin{proof}
For any point $x$ on any optimal path $\pi$, we have
the following equalities ($\lvert.\rvert$ denotes the link distance):

$\lvert\pi\rvert = D_\pi(s,x)+D_\pi(x,t) \hspace{13mm} or \hspace{12mm} \lvert\pi\rvert = D_\pi(s,x)+D_\pi(x,t)-1$

The first equation occurs, if $x$ is a bending point on $\pi$. In this case,
$D_\pi(s,x) = D_L(s,x)$ and $D_\pi(x,t) = D_L(x,t)$
due to the local optimality principle.
Similarly, the second equation takes place, if the last bending point from $s$ to $x$, $x$, and the first bending point from $x$ to $t$ are collinear.

Since two points $s$ and $t$ lie on the same side of $e$ inside $P-p$, there is always a bending point
$B$ in $p$ on $\pi$. Indeed, there are at most two points $B$ and $B'$, because if we have more than two bending points in $p$, the link distance of $\pi$ (optimal path) can be shortened by at least one link due to 
the triangle inequality (see Figure~\ref{fig:Oneside}).

We construct a path $\pi'$ like this: for $x \in Ce(i)$,
append $\pi_L(s,x)$ to $\pi_L(x,t)$.
The following inequality clearly holds:

$s_i+t_i-1 = D_L(s,x)+D_L(x,t)-1 \leq \lvert\pi'\rvert \leq D_L(s,x)+D_L(x,t) = s_i+t_i$

Thus, we have: $\lvert\pi\rvert \leq \lvert\pi'\rvert \leq s_i+t_i$ for $\pi$.
According to the above equalities, if we have one bending point $B$, then $\lvert\pi\rvert = D_\pi(s,B)+D_\pi(B,t)$.
In this case, $B$ must be in $Ce(i)$, i.e., $\lvert\pi\rvert = s_i+t_i$ (Figure~\ref{fig:Oneside}$(a)$).
On the other hand, there are three options for two bending points $B$ and $B'$ (Figure~\ref{fig:Oneside}$(b)$):

\begin{enumerate}
\item Only $B \in Ce(i)$ or $B' \in Ce(i)$. Again in this case, $\lvert\pi\rvert = s_i+t_i$.
      More precisely, if $B \in Ce(i)$, then $B' \in Ce(j)$, where $s_i+t_i=s_j+t_j$.

\item $(B$ and $B') \in Ce(i)$, then $\lvert\pi\rvert = D_\pi(s,B)+1+D_\pi(B',t)=s_i+t_i+1$.
       
\item Neither $B$ nor $B'$ belongs to \emph{Cells} with minimum value (e.g., $Ce(i)$).
Thus, $B \in Ce(j)$ and $B' \in Ce(k)$, where
      $s_j+t_j=s_k+t_k>s_i+t_i$.

\end{enumerate}
For the last two options, $\pi$ would not be an optimal path.
Therefore, only the first case can occur. The above discussion indicates that $\lvert\pi\rvert=s_i+t_i-1$ would not be possible.
\end{proof}

\begin{figure}
	\centering
    \includegraphics[width=9.5cm,keepaspectratio=true]{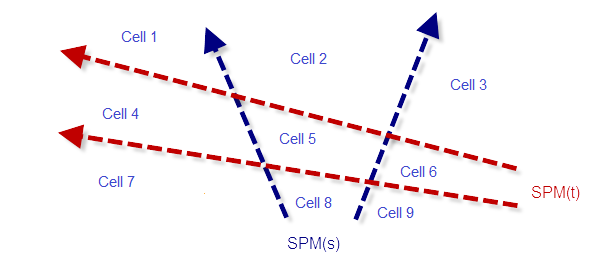}
	\caption{At most 4 intersections and 9 \emph{Cells}}
	\label{fig:Spms}
\end{figure}

To analyze the time complexity of the algorithm, observe that the check in Step $1$
can be easily done in linear time \cite {Suri_1986}.
The computations of $V(q)$ and $Pocket(q)$ in Step $2$ can be done using
the linear time algorithm in \cite {Joe_1987}.
Steps $3,4$ and $5$ can be done in linear time \cite {Edelsbrunner_1986,Suri_1986}.
Based on Lemma~\ref{Linear}, Step 6 can be done in linear time as well.

Note that, windows of a SPM never intersect each other.
According to Lemma~\ref{Intersection}, each window of $\mathrm{SPM(s)}$ intersects at most
two windows of $\mathrm{SPM(t)}$ and vice versa.
Therefore, if we have $k_1$ and $k_2$ windows of $\mathrm{SPM(s)}$ and $\mathrm{SPM(t)}$ inside $p$, respectively,
then there are at most $k_1+k_2$ intersection points inside $p$, where $k_1$, $k_2$ $=O(n)$.
Thus, we have $O(n)$ intersection points inside $p$ between the windows of $\mathrm{SPM(s)}$ and $\mathrm{SPM(t)}$.

To find the intersections, we make use of some known algorithms for subdivision
overlay like \cite {Finke_1995}, which solves the problem in optimal time
$O(n+k)$. Here, $k$ is the number of intersections, which is $O(n)$ in the worst case in our problem.
Therefore, Step $7$ can be done in linear time as well.

These intersections partition $p$ to at most $9/4(k_1+k_2)=O(n)$ \emph{Cells} (see Figure~\ref{fig:Spms}).
If either $k_1$ or $k_2$ does not exist, then we ignore the effect of that SPM to $p$. Furthermore,
if neither $k_1$ nor $k_2$ exists, the whole $p$ would be assumed as a \emph{Cell}.
Thus, Steps $8,9$ can be done in linear time.
According to \cite {Suri_1986}, Step $10$ is done in linear time.
The following theorem is concluded:

\begin{theorem}
For any three points $s$, $t$ and $q$ inside a simple polygon $P$ with total $n$ vertices, a $q$-$visible$ path
between $s$ and $t$ and its link distance can be computed in $O(n)$ time.
\end{theorem}

\section{Conclusion}
We studied the problem of finding a minimum link path between two points with point visibility constraint in a simple polygon.
The time bound for a $q$-$visible$ path is similar to the bound for the standard minimum link paths problem (without constraint). So, this bound cannot further be improved.

Possible extensions
to this problem involve studying the same problem when
the path is required to meet an arbitrary general region, not necessarily a visibility region
or having a query point instead of a fixed point.

A modified version of the problem is to constrain the path to meet several target polygons in a fixed order. If we fix the order of meeting, the problem seems to be less complex.
As Alsuwaiyel et al. \cite {Alsuwaiyel_1993} have shown, this problem is NP-hard even for any points on the boundary of a simple polygon.
This type of problem is defined as minimum link watchman route with no fixed order.

Another extension is to improve the approximation factor of minimum link visibility path
problem mentioned in \cite {Alsuwaiyel_1995},
using the same approach (SPM).

\section*{Acknowledgements}
The first author wishes to thank Dr Ali Gholami Rudi from Babol Noshirvani University of Technology
for many pleasant discussions.

\providecommand{\bysame}{\leavevmode\hbox
to3em{\hrulefill}\thinspace}

\end{document}